\documentclass[12pt,onecolumn,journal]{IEEEtran}
\usepackage{latexsym}
\usepackage{float}
\usepackage{amsfonts}
\usepackage{amsbsy}
\usepackage{amssymb}
\usepackage{times}
\usepackage{graphicx}
\usepackage{setspace}
\usepackage{enumerate}
\usepackage[usenames]{color}
\usepackage{epstopdf}
\usepackage[caption=false]{subfig}
\usepackage{cite}
\usepackage{amssymb}
\usepackage{amsfonts}
\usepackage{graphicx}
\usepackage{epsfig}
\usepackage{psfrag}
\usepackage{xcolor}
\usepackage{amsfonts, bm}
\usepackage{epstopdf}
\usepackage{cite}
\usepackage{color}
\usepackage{xcolor}
\usepackage{subfig}
\usepackage{verbatim}
\usepackage{multirow}
\usepackage{booktabs}
\usepackage{amsthm}
\usepackage{makecell}
\usepackage{units}
\usepackage[linesnumbered, ruled]{algorithm2e}
\usepackage{algpseudocode}
\usepackage{amsmath}
\usepackage{ulem}
\usepackage{colortbl}
\usepackage{color}

\IEEEoverridecommandlockouts
\usepackage{bibentry}

\begin{document}
	\title{Efficient Polarization Demosaicking via Low-cost Edge-aware and Inter-channel Correlation}
\author{Guangsen Liu, Peng Rao*, Xin Chen, Yao Li and Haixin Jiang
	\thanks{Guangsen Liu, Peng Rao*, Xin Chen, Yao Li and Haixin Jiang are with Shanghai Institute of Technical Physics, Chinese Academy of Sciences, Shanghai 200083, China (e-mail: liuguangsen20@mails.ucas.ac.cn; Peng\_rao@mail.sitp.ac.cn; @zju.edu.cn; ylcai@zju.edu.cn; jianghaixin22@mails.ucas.ac.cn).
} }
\maketitle

\begin{abstract}
Efficient and high-fidelity polarization demosaicking is critical for industrial applications of the division of focal plane (DoFP) polarization imaging systems. However, existing methods have an unsatisfactory balance of speed, accuracy, and complexity. This study introduces a novel polarization demosaicking algorithm that interpolates within a three-stage basic demosaicking framework to obtain DoFP images. Our method incorporates a DoFP low-cost edge-aware technique (DLE) to guide the interpolation process. Furthermore, the inter-channel correlation is used to calibrate the initial estimate in the polarization difference domain. The proposed algorithm is available in both a lightweight and a full version, tailored to different application requirements. Experiments on simulated and real DoFP images demonstrate that our two methods have the highest interpolation accuracy and speed, respectively, and significantly enhance the visuals. Both versions efficiently process a 1024 $\times$ 1024 image on an AMD Ryzen 5600X CPU in 0.1402s and 0.2693s, respectively. Additionally, since our methods only involve computational processes within a 5 $\times$ 5 window, the potential for parallel acceleration on GPUs or FPGAs is highly feasible.
\end{abstract}

\section{Introduction}
Polarization, a fundamental property of light, reveals several characteristics such as surface roughness\cite{kechiche2020polarimetric}, three-dimensional normals\cite{li2021near}, and material composition\cite{usmani2021deep} of objects. Utilizing these properties, polarization imaging technology captures polarization information and has been widely applied across various fields, including autonomous driving\cite{li2021illumination}, target detection\cite{romano2012day}, three-dimensional reconstruction\cite{zhu2019depth}, material identification\cite{hu2016polarization}, de-fogging\cite{liang2020effective}, and biomedical imaging\cite{garcia2018bio}.

\begin{figure}[htbp]
    \centering
    \includegraphics[width=0.5\linewidth]{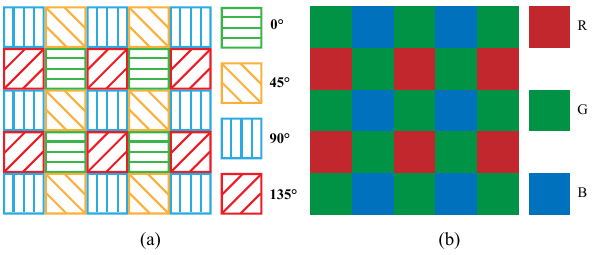}
    \caption{Schematic of the patterns of (a) a 5×5 PFA and (b) a 5×5 CFA.}
    \label{fig:PFA_CFA}
\end{figure}

Advancements in polarimetric imaging sensor technology have elevated the prominence of division of focal plane (DoFP) polarimetry in various polarimetric imaging systems, attributed to its miniaturization and real-time data acquisition capabilities. DoFP polarization imagers inherently compromise spatial resolution to facilitate the simultaneous acquisition of high-temporal-resolution optical signals from multiple polarization channels. DoFP cameras incorporate a polarization filter array (PFA) with orientations of \(0{}^\circ \), \(45{}^\circ \), \(90{}^\circ \), and \(135{}^\circ \) mirroring the pixel arrangement within the Color Filter Array (CFA), as illustrated in Fig.~\ref{fig:PFA_CFA}. Consequently, DoFP faces challenges akin to those encountered by CFA, often more complex in reconstructing the full polarization image from incomplete channel data. Such reconstruction necessitates that for each pixel of a DoFP image, the information from the remaining three channels must be accurately inferred to construct a complete description of the polarization state.

While color demosaicking has been extensively researched, yielding numerous efficient algorithms\cite{niu2018low,wu2016bayer,yang2020mcfd,lien2017efficient,menon2006demosaicing,kiku2016beyond}, these methods cannot be directly applied to polarization demosaicking due to distinct channel distributions and correlations\cite{mihoubi2018survey}.

Numerous interpolation-based algorithms have been developed to demosaic DoFP images, aiming to mitigate instantaneous field of view (IFoV) errors\cite{ratliff2009interpolation} and enhance spatial resolution. Traditional methods such as nearest-neighbor (NN), bilinear (BI), and bicubic (BCB)\cite{gao2011bilinear} perform interpolations on individual channels. As low-pass filters, these techniques often generate pronounced artifacts in high-frequency details. Zhang et al.\cite{zhang2016image} introduced an algorithm employing intensity correlation among polarization channels (ICPC) that identifies edges and interpolates images along a single direction; however, this binary decision approach is prone to significant information loss. Li et al.\cite{li2019demosaicking} proposed Newton's polynomial (NP) interpolation method to diminish artifacts and reduce information loss through weighted interpolations within the polarization difference domain. Nonetheless, this method requires a manually set, fixed decision threshold, leading to potential over- or under-estimation of pixel values. Morimatsu et al.\cite{morimatsu2020monochrome} developed the edge-aware residual interpolation (EARI) technique, which utilizes the intensity map as a base for guided filtering in the residual domain, enhancing interpolation precision, albeit at a significantly higher cost. Algorithms based on inter-channel correlation, such as PCDP\cite{wu2021polarization}, maintain low computational overhead by retaining only convolutional operations; nonetheless, they neglect directional interpolation, leading to visual errors—such as the ``zip'' effect—and mediocre performance on metrics such as PSNR, RMSE, and SSIM. PDEC\cite{xin2023demosaicking} addresses this by applying a single-channel Hamilton-Adam (HA) interpolation on bilinear-interpolated images. However, this merely blurs IFoV errors without genuinely resolving the loss of information during the interpolation process. These challenges highlight the importance of selecting appropriate interpolation directions for reconstructing high-precision polarization images. Despite extensive research, existing interpolation-based DoFP demosaicking methods still suffer from high computational costs and insufficient image quality in real-time applications.

Furthermore, optimization-based and data-driven algorithms have emerged as focal points of recent research. The sparse representation method\cite{zhang2018sparse} delivers considerable demosaicking results; however, its computational complexity limits its applicability to real-time imaging. Additionally, deep learning-based demosaicking\cite{li2023joint,zeng2019end,pistellato2022deep} has advanced; nonetheless, training such models requires high-quality data and substantial computational resources, and simultaneously acquiring DoFP and ground-truth images remains challenging. These issues constrain the broader industrial adoption of deep learning techniques for DoFP demosaicking.

In summary, although current demosaicking methods strive for simplicity, accuracy, and real-time execution, they still face challenges in balancing complexity, precision, and speed effectively. 

This study aims to develop a demosaicking method that reduces computational overhead while enhancing image quality, particularly for industrial applications. Our main contributions are as follows:

\begin{itemize}
	\item \textbf{TIPDF Framework:} We introduce a three-stage interpolation-based polarization demosaicking framework (TIPDF) to standardize the interpolation steps for four-channel DoFP images. This framework is divided into three primary phases: estimating orthogonal channel planes, determining non-orthogonal channel planes, and refining interpolation results.
	\item \textbf{LEIC and LEPD Interpolation:} Based on TIPDF, we develop a novel polarization demosaicking algorithm that employs a low-cost edge-aware and inter-channel correlation (LEIC) method to facilitate rapid high-quality reconstruction of DoFP polarization images. Our algorithm integrates a DoFP low-cost edge-aware (DLE) technique to render nimble interpolation decisions and an inter-channel correlation calibration (ICCC) to refine the initial estimates. For applications, a lighter version of LEIC, low-cost edge-aware polarization demosaicking (LEPD), has also been introduced. We present comprehensive quantitative assessments and visual comparisons of our reconstruction with other established algorithms using publicly available datasets.
	\item \textbf{OLVD Dataset:} We create an outdoor long-wave infrared and visible DoFP polarization image dataset (OLVD) which includes thousands of LWIR and VIS DoFP images captured simultaneously or individually. Several scenes of OLVD are used to evaluate the effectiveness of different demosaicking methods on real DoFP images. The dataset will provide a valuable contribution to subsequent research.
\end{itemize}

\section{Linear polarization calculation}
The DoFP detector acquires polarization images by capturing light through four distinct polarization filters. The polarization state of light can be represented by a Stokes vector, denoted as \(S = \left[ S_{0}, S_{1}, S_{2}, S_{3} \right]\). The first three components of $S$ are defined as follows\cite{pistellato2022deep}:
\begin{equation}
    S_{0}=0.5(I_{0} + I_{45} + I_{90} + I_{135}), \tag{1a}
\end{equation}
\begin{equation}
    S_{1}=I_{0} - I_{90}, \tag{1b}
\end{equation}
\begin{equation}
    S_{2}=I_{45} - I_{135}, \tag{1c}
\end{equation}
where \(I_{\theta}\) represents the intensities measured by the DoFP detector at polarization angles of $\theta$; \(S_{0}\) corresponds to the total intensity of incident light, whereas \(S_{1}\) and \(S_{2}\) correspond to the linear polarization components. The Degree of Linear Polarization (\(\textit{DoLP}\)) and the Angle of Linear Polarization (\(\textit{AoLP}\)) are commonly used to describe polarization attributes, which are expressed as 
\begin{equation}
    \textit{DoLP}=\frac{\sqrt{S_{1}^2 + S_{2}^2}}{S_{0}}, \tag{2a}
\end{equation}
\begin{equation}
    \textit{AoLP}=0.5\arctan\left(\frac{S_{2}}{S_{1}}\right). \tag{2b}
\end{equation}

\begin{figure}[htbp]
    \centering
    \includegraphics[width=0.5\linewidth]{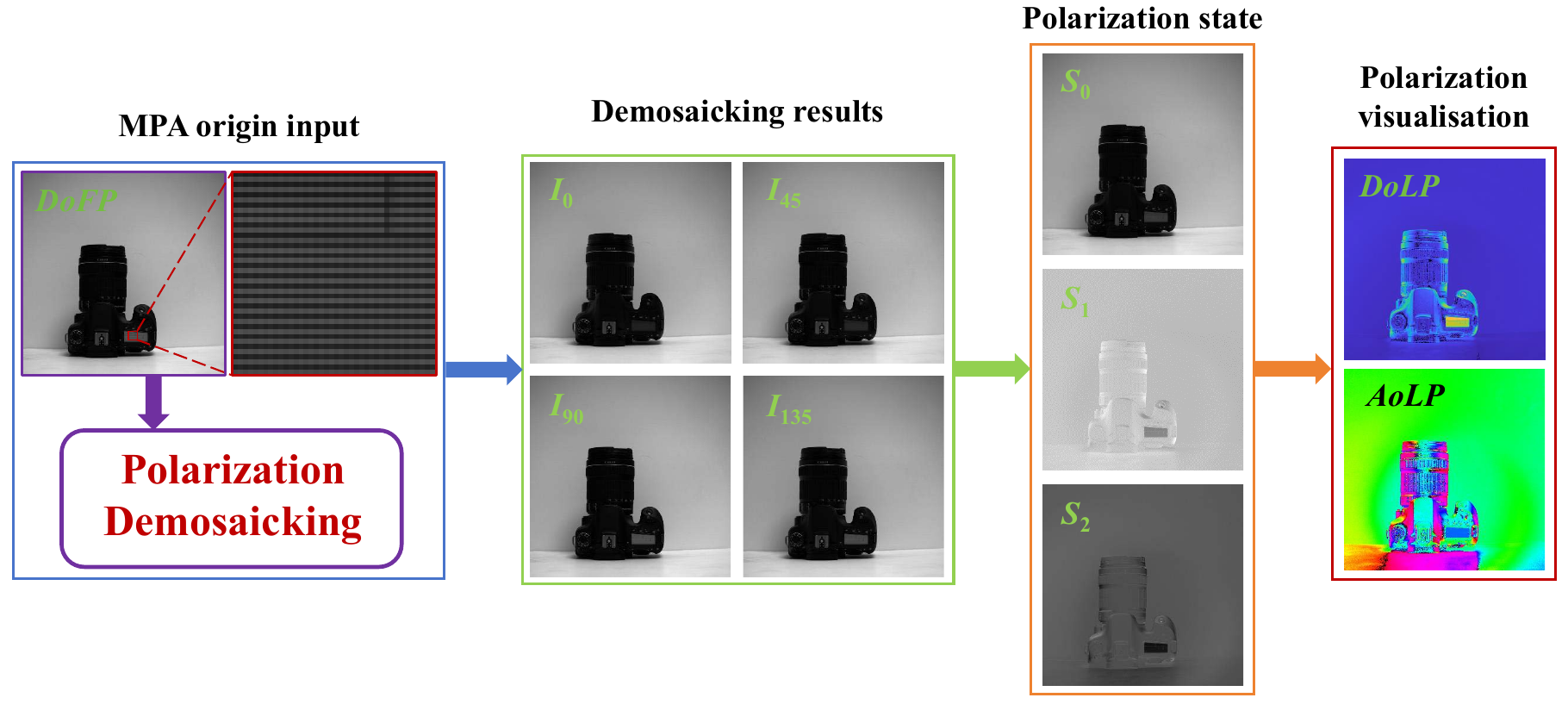}
    \caption{Reconstruction workflow for DoFP polarization images.}
    \label{fig:reconstruction_workflow}
\end{figure}

The process of reconstructing and visualizing polarization images is illustrated in
Fig.~\ref{fig:reconstruction_workflow}.

\section{Proposed Methods}

\begin{figure} [htbp]
    \centering
    \includegraphics[width=0.6\linewidth]{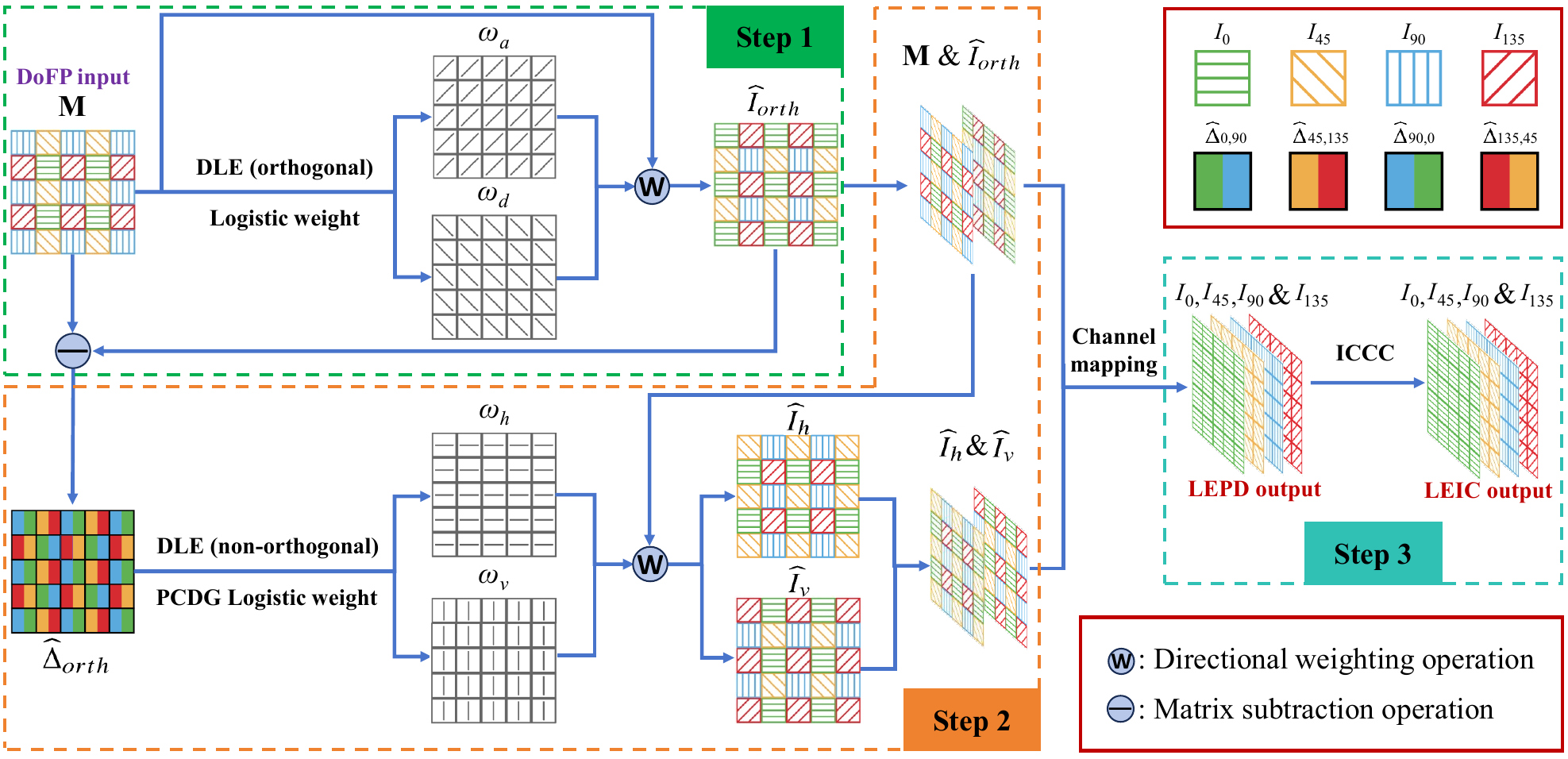}
    \caption{Flowchart illustrating our polarization demosaicking process. Steps 1 to 3 each correspond to the three stages of TIPDF.}
    \label{fig:LEIC}
\end{figure}

This section elaborates on the technical details of the demosaicking algorithm developed in this study. Initially, we introduce a TIPDF to guide the demosaicking process. Subsequently, detailed computations are provided for the DLE technique, orthogonal and non-orthogonal channel plane estimation, and ICCC. A flowchart illustrating our polarization demosaicking process is depicted in Fig.~\ref{fig:LEIC}.

\subsection{Polarization demosaicking framework}

The TIPDF is proposed to standardize the DoFP demosaicking operation, as illustrated in Fig.~\ref{fig:LEIC}. This framework is inspired by the highly successful CFA image demosaicking framework. In color demosaicking, the process typically begins with the interpolation of the green channel, followed by the estimation of the remaining channels, and concludes with corrections based on the observed pixels. 

For DoFP images, the orthogonal channel pixels are arranged centrosymmetrically among the neighboring pixels of the interpolated pixel, facilitating directional interpolation. The orthogonal channel also occupies a significant proportion of the neighborhood, which supports the initial focus on these channels. Therefore, the missing orthogonal channel values corresponding to each pixel are estimated first, ensuring that half of the pixels in each channel plane have values. Subsequently, both the DoFP and the orthogonal channel plane are utilized to estimate the remaining two channels, compensating for the missing pixels across all channels. The interpolation results inevitably contain certain errors, and the observed pixels of the DoFP image represent the only accurate pixel information available. To maximize their effectiveness, inter-channel correlation\cite{wu2021polarization,xin2023demosaicking} is employed to further calibrate the polarization image, enhancing the overall accuracy of the demosaicking process.

\subsection{DoFP low-cost edge-aware}
The Hamilton–Adams demosaicking algorithm (HA) is widely recognized for demosaicking CFA images due to its ability to achieve notable results through simple computations\cite{hamilton1997adaptive,buades2009self,gharbi2016deep}. We have adapted the HA algorithm for PFA images to streamline the interpolation decision-making process for DoFP images. The application of the HA algorithm in orthogonal channels is detailed below.

Consider $\mathbf{M}$, the original DoFP image, consisting of $m$ rows and $n$ columns. The set of all pixel coordinates is defined as follows:
\[L=\{(i,j)\in {{\mathbb{N}}^{2}}|i\in [1,m],j\in [1,n]\}.\tag{3}\]

Firstly, the computations of first and second-order diagonal and anti-diagonal partial derivatives of image $\mathbf{M}$ are conducted as follows:
\[{{\partial }_{d}}\mathbf{M}(i,j)=\frac{\mathbf{M}(i+1,j+1)-\mathbf{M}(i-1,j-1)}{2\sqrt{2}},\tag{4a}\]
\[\partial _{d}^{2}\mathbf{M}(i,j)=\frac{\mathbf{M}(i+2,j+2)+\mathbf{M}(i-2,j-2)-2\mathbf{M}(i,j)}{8},\tag{4b}\]
\[{{\partial }_{a}}\mathbf{M}(i,j)=\frac{\mathbf{M}(i-1,j+1)-\mathbf{M}(i+1,j-1)}{2\sqrt{2}},\tag{4c}\]
\[\partial _{a}^{2}\mathbf{M}(i,j)=\frac{\mathbf{M}(i-2,j+2)+\mathbf{M}(i+2,j-2)-2\mathbf{M}(i,j)}{8},\tag{4d}\]
where $d$ and $a$ denote the diagonal and anti-diagonal directions, respectively.

Next, we calculate the local intensity variations as follows:
\[{{v}_{d}}=\left| {{\partial }_{d}}\mathbf{M}(i,j) \right|+\left| 2\sqrt{2}\partial _{d}^{2}\mathbf{M}(i,j) \right|,\tag{5a}\]
\[{{v}_{a}}=\left| {{\partial }_{a}}\mathbf{M}(i,j) \right|+\left| 2\sqrt{2}\partial _{a}^{2}\mathbf{M}(i,j) \right|.\tag{5b}\]
The average of the neighboring orthogonal channels in the diagonal and anti-diagonal directions are expressed as
\[\overline{I}_{orth}^{d}=\frac{1}{2}(\mathbf{M}(i+1,j+1)+\mathbf{M}(i-1,j-1)),\tag{6a}\]
\[\overline{I}_{orth}^{a}=\frac{1}{2}(\mathbf{M}(i+1,j-1)+\mathbf{M}(i-1,j+1)).\tag{6b}\]

Further, the estimation of the orthogonal channel value at a given position $(i, j)$ is expressed by
\[{{\widehat{I}}_{orth}}(i,j)={{\omega }_{d}}(\overline{I}_{orth}^{d}-\partial _{d}^{2}\mathbf{M}(i,j))+{{\omega }_{a}}(\overline{I}_{orth}^{a}-\partial _{a}^{2}\mathbf{M}(i,j)),\tag{7}\]
where \({{\omega }_{d}}\) and \({{\omega }_{a}}\) denote the interpolation weights for the diagonal and anti-diagonal directions, respectively. These weights are determined according to variations \({{v}_{d}}\) and \({{v}_{a}}\), and they satisfy the normalization condition \({{\omega }_{d}}+{{\omega }_{a}}=1\).

\subsection{Orthogonal channel plane estimation}
The DLE is initially employed to estimate the values of the orthogonal channels, thereby constructing the orthogonal channel plane. In certain previous demosaicking methods, edge detection primarily relied on a basic ternary edge classifier\cite{hamilton1997adaptive,wu2016bayer}:

\begin{equation}
{{\omega }_{d}}=\begin{cases}
0 & \text{if }{{v}_{d}}-{{v}_{a}}<-T \\ 
1 & \text{if }{{v}_{d}}-{{v}_{a}}>T \\ 
0.5 & \text{if }\left| {{v}_{d}}-{{v}_{a}} \right|\le T \\ 
\end{cases}.\tag{8}
\end{equation}
However, a limitation of this approach is the selection of the threshold $T$ for different scenarios. An ill-selected $T$ can lead to opposing estimation results, which is undesirable. Therefore, a continuous function $f(v_{d}-v_{a})$ is used to represent $\omega_{d}$. In this study, we use the logistic function to calculate $\omega_{d}$ as follows:
\[\omega_{d} = f(v_{d} - v_{a}) = \frac{1}{1 + e^{k(v_{d} - v_{a})}},\tag{9a}\]

\[k = k_{0} \frac{\textit{DR}(\mathbf{M})}{255},\tag{9b}\]
where $k$ is a parameter used to modify the change in weights. The logistic function ensures a continuous adjustment of the weights, preventing the incorrect decisions that can occur with a step function. Furthermore, $\omega_{a} = f(v_{a} - v_{d}) = 1-\omega_{d}$, an equivalent form of calculation for both weights. Given the inverse relationship between $k$ and the range of $(v_{d} - v_{a})$, and a direct relationship between $(v_{d} - v_{a})$ and the dynamic range of the DoFP image $\textit{DR}(\mathbf{M})$, $k$ is proportional to $\textit{DR}(\mathbf{M})$. The value of $k_{0}$ represents $k$ when $\textit{DR}(\mathbf{M}) = 255$. Experimental results using a range of $k_{0}$ values ranging from 0.1 to 10.0 with a step size of 0.1 on Qiu et al.'s dataset\cite{qiu2019polarization} revealed that the optimal PSNR was obtained at $k_{0} = 1.0$. Thus, in this study, $k_{0} = 1.0$ has been maintained.

Eq. (7) utilizes the weights formulated in Eq. (9) to estimate the orthogonal channel plane $\widehat{I}_{\textit{orth}}$. The continuity of the logistic function will help the DLE to make flexible interpolation decisions.

\subsection{Non-orthogonal channel plane estimation}
In a DoFP image, each pixel's horizontally and vertically neighboring pixels correspond to different polarization channels. Relying solely on the observed values for estimation inevitably discards valuable information from the other direction. The orthogonal channel plane addresses this information loss, compensating for it by utilizing the fact that the horizontally and vertically neighboring pixels are channels orthogonal to each other. 

\begin{figure} [htbp]
    \centering
    \includegraphics[width=0.6\linewidth]{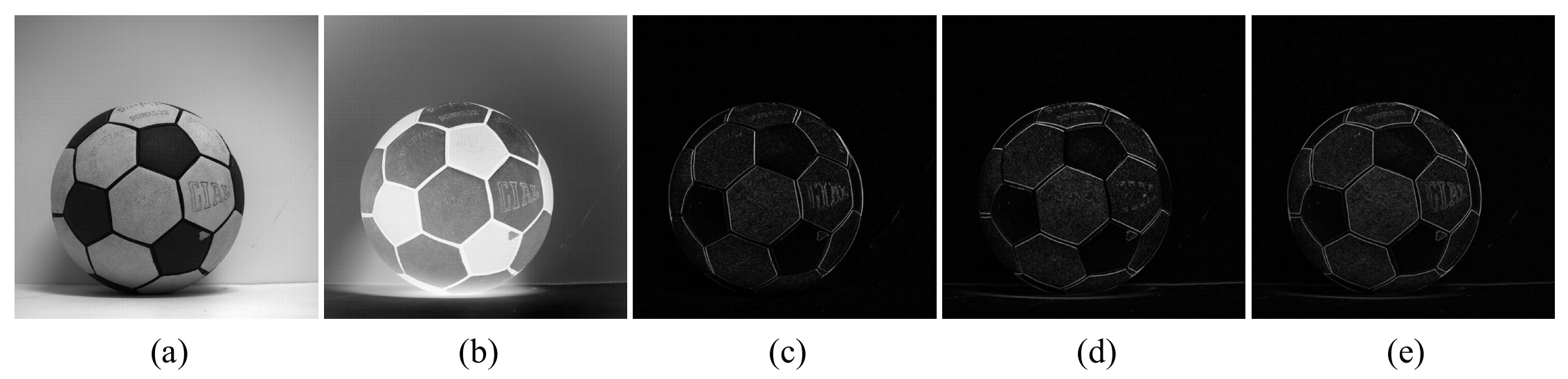}
    \caption{ Visualization of polarization channel difference gradients (PCDG). (a) DoFP image. (b) Visualization of orthogonal polarization channel difference. (c) PCDG in the horizontal direction. (d) PCDG in the vertical direction (e) Absolute PCDG.}
    \label{fig:PCDG}
\end{figure}

In CFA demosaicking, chrominance gradients are more pronounced in the direction perpendicular to edges rather than parallel to them\cite{sun2019hybrid}. We adopt this perspective for DoFP images, suggesting that the polarization channel difference gradient (PCDG) is more significant in the trans-edge direction, as visually depicted in Fig.~\ref{fig:PCDG}. In this study, PCDG serves as a reliable parameter for reflecting edges and is utilized for edge-aware processing in horizontal and vertical channels. 

Firstly, the orthogonal channel difference map is formulated as
\[{{\widehat{\Delta }}_{orth}}(i,j)=\mathbf{M}(i,j)-{{\widehat{I}}_{orth}}(i,j).\tag{10}\]

The first and second-order horizontal and vertical partial derivatives of the orthogonal channel residual map are computed as
\[{{\partial }_{h}}{{\widehat{\Delta }}_{orth}}(i,j)=\frac{{{\widehat{\Delta }}_{orth}}(i,j+1)-{{\widehat{\Delta }}_{orth}}(i,j-1)}{2},\tag{11a}\]
\[\partial _{h}^{2}{{\widehat{\Delta }}_{orth}}(i,j)=\frac{{{\widehat{\Delta }}_{orth}}(i,j+2)+{{\widehat{\Delta }}_{orth}}(i,j-2)-2{{\widehat{\Delta }}_{orth}}(i,j)}{4},\tag{11b}\]
\[{{\partial }_{v}}{{\widehat{\Delta }}_{orth}}(i,j)=\frac{{{\widehat{\Delta }}_{orth}}(i+1,j)-{{\widehat{\Delta }}_{orth}}(i-1,j)}{2},\tag{11c}\]
\[\partial _{v}^{2}{{\widehat{\Delta }}_{orth}}(i,j)=\frac{{{\widehat{\Delta }}_{orth}}(i+2,j)+{{\widehat{\Delta }}_{orth}}(i-2,j)-2{{\widehat{\Delta }}_{orth}}(i,j)}{4},\tag{11d}\]
where $h$ and $v$ represent the horizontal and vertical directions, respectively.

Next, PCDGs related to edge variations in the horizontal and vertical directions are computed as follows:
\[{{v}_{h}}=\left| {{\partial }_{h}}{{\widehat{\Delta }}_{orth}}(i,j) \right|+\left| 2\partial _{h}^{2}{{\widehat{\Delta }}_{orth}}(i,j) \right|,\tag{12a}\]
\[{{v}_{v}}=\left| {{\partial }_{v}}{{\widehat{\Delta }}_{orth}}(i,j) \right|+\left| 2\partial _{v}^{2}{{\widehat{\Delta }}_{orth}}(i,j) \right|.\tag{12b}\]

Subsequently, the average of the neighboring horizontal and vertical channels in the horizontal and vertical directions is calculated from the DoFP plane and orthogonal channel plane by
\[\overline{I}_{h}^{h}=\frac{1}{2}(\mathbf{M}(i,j+1)+\mathbf{M}(i,j-1)),\tag{13a}\]
\[\overline{I}_{h}^{v}=\frac{1}{2}({{\widehat{I}}_{orth}}(i+1,j)+{{\widehat{I}}_{orth}}(i-1,j)),\tag{13b}\]
\[\overline{I}_{v}^{h}=\frac{1}{2}({{\widehat{I}}_{orth}}(i,j+1)+{{\widehat{I}}_{orth}}(i,j-1)),\tag{13c}\]
\[\overline{I}_{v}^{v}=\frac{1}{2}(\mathbf{M}(i+1,j)+\mathbf{M}(i-1,j)),\tag{13d}\]
where the subscript of \(\overline{I}\) represents the channel, $h$ indicates horizontal neighbor and $v$ denotes vertical neighbor with respect to the interpolated pixel; the superscript refers to the direction of interpolation.

Similar to Eq. (4b) and (4d), the second-order horizontal and vertical partial derivatives of DoFP are computed as
\[\partial _{h}^{2}\mathbf{M}(i,j)=\frac{\mathbf{M}(i,j+2)+\mathbf{M}(i,j-2)-2\mathbf{M}(i,j)}{4},\tag{14a}\]	
\[\partial _{v}^{2}\mathbf{M}(i,j)=\frac{\mathbf{M}(i+2,j)+\mathbf{M}(i-2,j)-2\mathbf{M}(i,j)}{4}.\tag{14b}\]	

Finally, the value of horizontal and vertical channels at position \(\left( i,j \right)\) are estimated by
\[{{\widehat{I}}_{h}}(i,j)={{\omega }_{h}}(\overline{I}_{h}^{h}-\partial _{h}^{2}\mathbf{M}(i,j))+{{\omega }_{v}}(\overline{I}_{h}^{v}-\partial _{v}^{2}\mathbf{M}(i,j)),\tag{15a}\]
\[{{\widehat{I}}_{v}}(i,j)={{\omega }_{h}}(\overline{I}_{v}^{h}-\partial _{h}^{2}\mathbf{M}(i,j))+{{\omega }_{v}}(\overline{I}_{v}^{v}-\partial _{v}^{2}\mathbf{M}(i,j)),\tag{15b}\]
\[{{\omega }_{h}}=f({{v}_{h}}-{{v}_{v}})=\frac{1}{1+{{e}^{k({{v}_{h}}-{{v}_{v}})}}},\tag{15c}\]
\[{{\omega }_{v}}=f({{v}_{v}}-{{v}_{h}})=\frac{1}{1+{{e}^{k({{v}_{v}}-{{v}_{h}})}}}=1-{{\omega }_{h}},\tag{15d}\]
where $\omega_h$ and $\omega_v$ denote the horizontal and vertical directional weights, respectively; $k$ is calculated by Eq. (9b), with parameter $k_0=1.0$, as introduced in Section 2.3.

At this point, we deduce estimates for the remaining three channels at the location \(\left( i,j \right)\):  \({{\widehat{I}}_{orth}}(i,j)\), \({{\widehat{I}}_{h}}(i,j)\) and \({{\widehat{I}}_{v}}(i,j)\).

These estimates are extended across the entire image and successively mapped onto each polarization channel plane to obtain the array of demosaicked images: \(~[{{\widehat{I}}_{0}}, {{\widehat{I}}_{45}}, {{\widehat{I}}_{90}}, {{\widehat{I}}_{135}}]\).

\subsection{Calibration with inter-channel correlation}
After the above process, we obtained the demosaicked images of the four channels. As each pixel's complementary, three channels are interpolated independently, and fidelity can be further enhanced through the use of the ICCC technique.

ICCC is based on a difference plane interpolation technique commonly employed in CFA demosaicking\cite{li2008image,chung2006color}. It leverages color correlation to significantly improve interpolation precision by focusing on the difference plane, which mitigates interpolation inaccuracies introduced by the image's high-frequency content. The efficacy of this method has been validated in DoFP demosaicking applications.

Given the strong correlation among polarization channels\cite{xin2023demosaicking}, calibrations through ICCC involve mutual referencing among the four pre-interpolated images.
The method initially estimates the mean local polarization channel differences and subsequently adjusts their interpolation to refine the overall image quality. The details are as follows:

For an individual channel $x$ $(x\in \left\{ 0,45,90,135 \right\})$, a sparse raw plane ${{\widetilde{\mathbf{M}}}_{x}}$ is computed by
\[{{\widetilde{\mathbf{M}}}_{x}}=\mathbf{M}\odot {\textit{mask}_{x}},\tag{16}\]
where $\odot $ symbolizes the Hadamard product and ${\textit{mask}_{x}}$ is a binary matrix at the coordinate \(\left( i,j \right)\), defined by
\begin{equation}
\textit{mask}_x(i,j)=\begin{cases}
1, & \text{if } (i,j)\in P_x,\\
0, & \text{otherwise},
\end{cases}\tag{17}
\end{equation}
where \({{P}_{x}}\) represents the subset of $x$-channel.

The correction of each channel estimate is performed individually.
Considering the \(0{}^\circ\) channel as an example, we initially compute the sparse differential estimation plane relative to the \(0{}^\circ\) channel for the other three channels at the positions $P_0$ as follows:

\[{{\widetilde{\Delta }}_{c,0}}={{\widetilde{\mathbf{M}}}_{0}}-{{\widehat{I}}_{c}}\odot {\textit{mask}_{0}}, \quad c\in \left\{45,90,135 \right\}.\tag{18}\]

Under the assumption of neighbor channel difference consistency\cite{wu2021polarization}, bilinear interpolation is used to obtain three complete difference planes:
\[{{\widehat{\Delta }}_{c,0}}={{\widetilde{\Delta }}_{c,0}}*F, \quad c\in \left\{45,90,135 \right\}.\tag{19}\]

Next, the correction of each channel to \(0{}^\circ \) channel is obtained by summing the estimation planes with the difference planes of the corresponding channels:

\[{{\widehat{I}}_{c,0}}={{\widehat{I}}_{c}}+{{\widehat{\Delta }}_{c,0}}, \quad c\in \left\{45,90,135 \right\}.\tag{20}\]

The final refinement for the $0^\circ$ channel is achieved through the weighted mean of the three individual channel corrections mentioned above:
\[{{I}_{0}}=\sum\limits_{c}{{{\omega }_{c}}{{\widehat{I}}_{c,0}}},\tag{21}\]
where $c$ denotes the other three channels, i.e., \(c\ne 0\). \({{\omega }_{c}}\) denotes the weight of the $c$-channel, with \(\sum\limits_{c}{{{\omega }_{c}}}=1\). Since horizontal and vertical channels have an equal and more substantial influence than orthogonal channels, Eq. (21) can be simplified as follows:
\[{{I}_{0}}={{\omega }_{hv}}({{\widehat{I}}_{45,0}}+{{\widehat{I}}_{135,0}})+{{\omega }_{orth}}{{\widehat{I}}_{90,0}},\tag{22}\]
where \({{\omega }_{hv}}\) denotes the horizontal and vertical channel weights, and \({{\omega }_{orth}}\) denotes the orthogonal channel weights with \(2{{\omega }_{hv}}+{{\omega }_{orth}}=1\). The values of these weights correlate with their contribution to the final image, with a theoretical bias, such that, \({{\omega }_{hv}}>{{\omega }_{orth}}\). The global weights are deduced from the polarization distance\cite{wu2021polarization} as follows:

\[
\begin{cases}
  \omega_{hv} = \frac{\sqrt{2}}{1+2\sqrt{2}}, \\ 
  \omega_{orth} = \frac{1}{1+2\sqrt{2}}.
\end{cases}
\tag{23}
\]

The outlined ICCC method can be easily extended to the other three channels to obtain all four calibrated images. Therefore, a detailed exposition of this procedure is omitted here.

The introduction of ICCC significantly improves the interpolation accuracy and effectively eliminates redundant textures, as will be discussed in detail in Section 3.

\subsection{Polarization demosaicking using LEIC and LEPD}

Finally, we integrate the computational steps described above to introduce a novel polarization demosaicking algorithm, LEIC. The algorithm flow and a visual representation of LEIC, along with its lightweight version, LEPD, are depicted in Algorithm~\ref{alg:r2p} and Fig.~\ref{fig:LEIC}, respectively.

\begin{algorithm} [htbp]
    \caption{Proposed LEIC Method}  
    \label{alg:r2p}  
    \KwIn{DoFP raw image; Version (LEPD or LEIC)}  
    \KwOut{Demosaicked images \({{I}_{0}}\), \({{I}_{45}}\), \({{I}_{90}}\), and \({{I}_{135}}\)
    }
    \textbf{// Step 1. Estimate Orthogonal Channel Plane by DLE}\\
    Compute the diagonal and anti-diagonal weights, \({{\omega }_{d}}\) and \({{\omega }_{a}}\), by Eq. (4a) to (4d), (5a), (5b), (9a), and (9b).\\
    Estimate the value of the orthogonal channel plane \({{\widehat{I}}_{orth}}\) using Eq. (4b), (4d), (6a), (6b), (7), \({{\omega }_{d}}\), and \({{\omega }_{a}}\).\\
    \textbf{// Step 2. Estimate Non-orthogonal Channel Plane by DLE}\\
    Calculate the orthogonal channel difference map \({{\widehat{\Delta }}_{orth}}\) by Eq. (10).\\
    Calculate the horizontal and vertical weights, \({{\omega }_{h}}\) and \({{\omega }_{v}}\), by Eq. (11a)$ \sim$(11d), (12a), (12b), (15c), and (15d).\\
    Estimate the value of horizontal and vertical channel plane, \({{\widehat{I}}_{h}}\) and \({{\widehat{I}}_{v}}\), using Eq. (13a) to (13d), (14a), (14b), (15a), (15b), \({{\omega }_{h}}\), and \({{\omega }_{v}}\).\\
    Generate 4 demosaicked images (\({{I}_{0}}\), \({{I}_{45}}\), \({{I}_{90}}\), and \({{I}_{135}}\)) by extracting corresponding values from \(\mathbf{M}\), \({{\widehat{I}}_{orth}}\), \({{\widehat{I}}_{h}}\), and \({{\widehat{I}}_{v}}\).\\
        \If {version is LEPD}
        {\textbf{return} \({{I}_{0}}\), \({{I}_{45}}\), \({{I}_{90}}\), and \({{I}_{135}}\). \textbf{(The lightweight version: LEPD)}}
    \textbf{// Step 3. Update with Observed Data by ICCC}\\
        \ForEach {channel $x$}{
            Compute the difference plane of the other 3 channels \({{\widehat{\Delta }}_{c,x}}\) using Eq. (16) to (19);
            
            Compute the corrections for the other 3 channels \({{\widehat{I}}_{c,x}}\) using Eq. (20);
            
            Refine the demosaicked images for channel $x$, \({{I}_{x}}\), using Eq. (22) with global channel weights \({{\omega }_{hv}}\) and \({{\omega }_{orth}}\) as defined in Eq. (23).}
    {\textbf{return} {\({{I}_{0}}\), \({{I}_{45}}\), \({{I}_{90}}\), and \({{I}_{135}}\). \textbf{(The full version: LEIC)}}}
\end{algorithm}





\section{ Experiments}

In this section, we design three experiments to assess the performance of our LEPD and LEIC algorithms. We compare the quantitative metrics, visual effects, and running speeds of different interpolation methods using simulated DoFP images. Subsequently, the methods were tested on a self-built real DoFP images dataset, the OLVD. Traditional methods have been used for comparison, such as NN, BI, and BCB\cite{gao2011bilinear}, and advanced methods such as ICPC\cite{zhang2016image}, EARI\cite{morimatsu2020monochrome}, PCDP\cite{wu2021polarization}, PDEC\cite{xin2023demosaicking}, NP\cite{li2019demosaicking}, and our LEPD and LEIC have been employed. 
Technical descriptions of these methods are listed in Table~\ref{tab:methods}. 

\begin{table}[ht!]
    \small
    \centering
    \caption{\textbf{Technical Descriptions of 10 Polarization Demosaicking Methods}}
    \begin{tabular}{c | c | c | c}
    \toprule[2pt]
        \textbf{Method} & \textbf{Interpolation Decision} & \textbf{Initial Estimation} & \textbf{Calibration} \\
        \midrule[1pt]
        \textbf{NN} & No & Nearest-Neighbor Interpolation & No \\
        \textbf{BI} & No & Bilinear Interpolation & No \\
        \textbf{BCB} & No & Bicubic Interpolation & No \\
        \textbf{ICPC} & Edge Binary Decision & Bicubic Spline Interpolation & No \\
        \textbf{EARI} & Directional-Conv Edge-aware & Residual Interpolation & No \\
        \textbf{PCDP} & No & Bilinear Interpolation & Inter-Channel Correlation \\
        \textbf{PDEC} & No & Bilinear \& Edge Compensation & Inter-Channel Correlation \\
        \textbf{NP} & Weighted Edge Classifier & Newton’s Polynomial Interpolation & No \\
        \textbf{\uline{LEPD}} & DoFP Low-cost Edge-aware & Hamilton-Adam Interpolation & No \\
        \textbf{\uline{LEIC}} & DoFP Low-cost Edge-aware & Hamilton-Adam Interpolation & Inter-Channel Correlation \\
    \bottomrule[2pt]
    \end{tabular}
    
    \label{tab:methods}
\end{table}

\subsection{Performance assessment on simulated images}

\begin{figure}[htbp]
    \centering
    \includegraphics[width=0.5\linewidth]{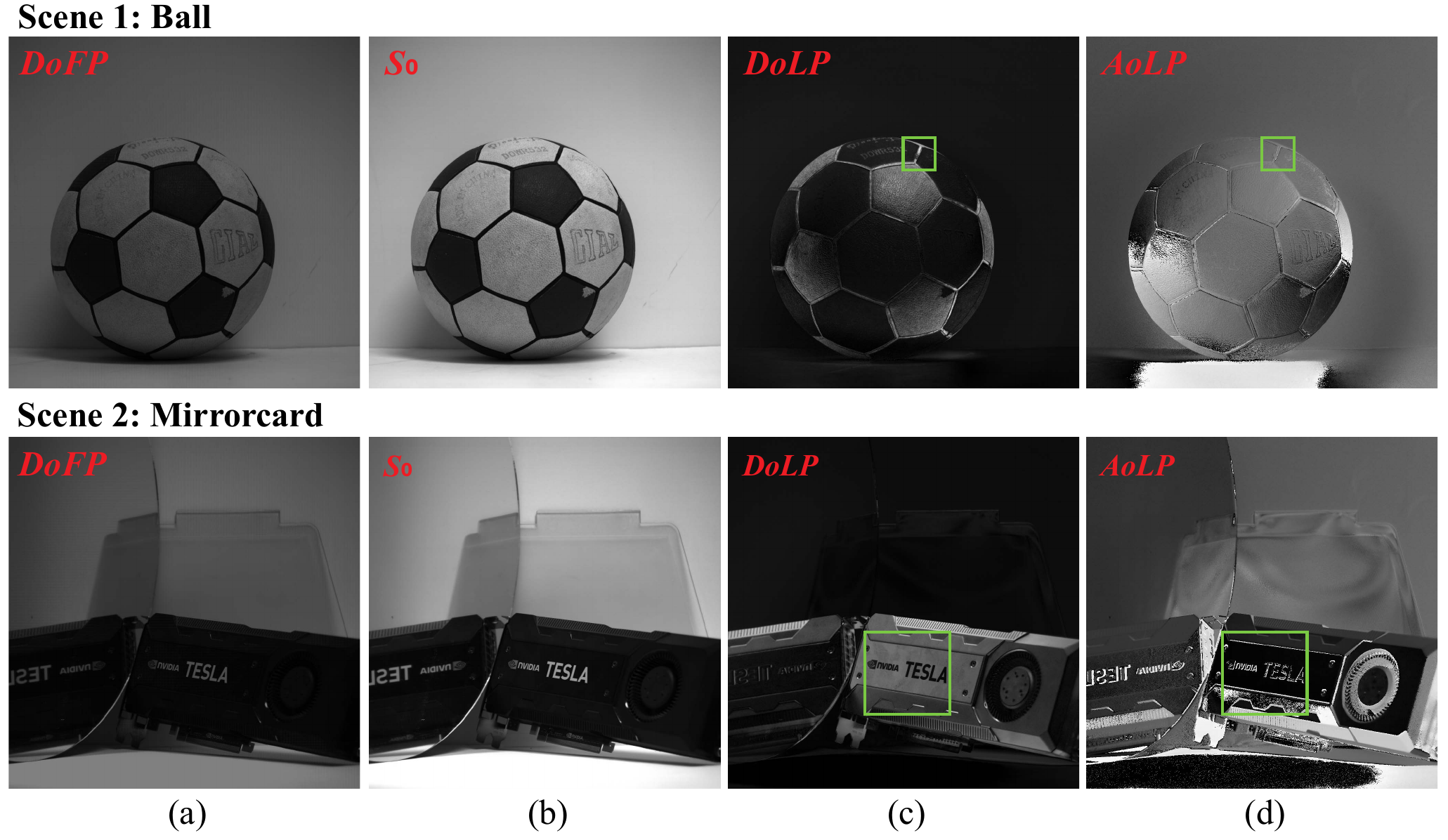}
    \caption{Selected scenes (Ball and Mirrorcard) from Qiu's dataset~\cite{qiu2019polarization} applied in the test experiments. (a) Simulated DoFP images; (b) \({{S}_{0}}\) images; (c) $\textit{DoLP}$ images; (d) $\textit{AoLP}$ images.}
    \label{fig:simulated_dofp_scenes}
\end{figure}

\begin{figure}[htbp]
    \centering
    \includegraphics[width=0.5\linewidth]{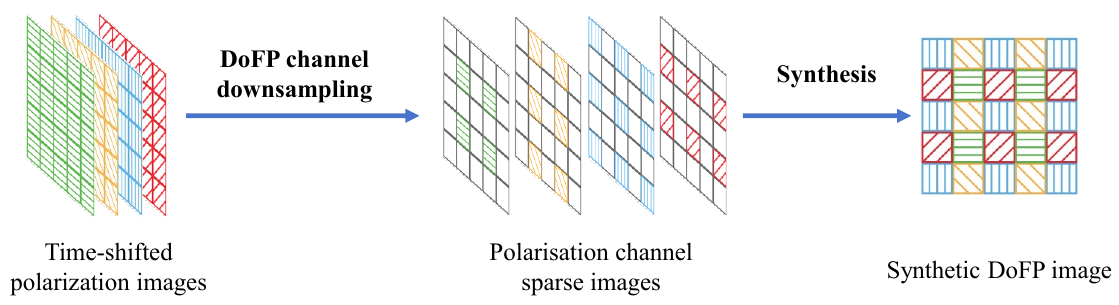}
    \caption{Simulation process of DoFP images using time-shifted polarization images.}
    \label{fig:synthesis_process}
\end{figure}

Using the time-shifted polarization image dataset provided by Qiu et al.~\cite{qiu2019polarization}, we synthesized simulated DoFP images with a resolution of 1024 $\times$ 1024, which are presented in Fig.~\ref{fig:simulated_dofp_scenes}. As illustrated in Fig.~\ref{fig:synthesis_process}, images from the four polarization channels at full resolution are downsampled according to the PFA pattern and subsequently fused to create a simulated DoFP image.

The simulated DoFP images from 40 scenes in the dataset were demosaicked using the 10 methods  previously mentioned. The average PSNR, RMSE, and SSIM results are listed in Tables~\ref{tab:psnr},~\ref{tab:rmse}, and~\ref{tab:ssim}, respectively. These metrics assess convey signal fidelity (PSNR), accuracy of value estimation (RMSE), and similarity between two images (SSIM). In each table, the top three results are highlighted in red, green, and blue to indicate the best, second-best, and third-best outcomes, respectively.

In these quantitative tests, our LEIC method consistently produces superior results across most metrics. The PDEC method exhibits improved RMSE and SSIM scores compared with LEIC but underperforms in other metrics. Overall, NP is ranked as the second-best method following LEIC; however, it demonstrates particularly inferior performance in reconstructed DoLP images across all metrics. LEIC significantly outperforms NP, which also employs edge-aware techniques. Additionally, the LEPD also performs well in most metrics, affirming the high efficacy of our reconstruction strategy even without the calibration process.

\begin{table}[htbp]
    \scriptsize
    \centering
    \caption{\textbf{Average PSNR on Simulated Images}}
    \begin{tabular}{c | c | c | c | c | c | c | c | c | c | c}
    \toprule[2pt]
        \textbf{PSNR} & \textbf{NN} & \textbf{BI} & \textbf{BCB} & \textbf{ICPC} & \textbf{EARI} & \textbf{PCDP} & \textbf{PDEC} & \textbf{NP} & \textbf{\uline{LEPD}} & \textbf{\uline{LEIC}} \\
        \midrule[1pt]
        \({\boldsymbol{I}}_{\boldsymbol{0}}\) & 36.502 & 41.049 & 42.022 & 41.768 & 41.604 & 43.128 & 43.245 & \textbf{\textcolor[rgb]{0, 0.7, 0}{43.664}} & \textbf{\textcolor[rgb]{0, 0, 0.7}{43.575}} & \textbf{\textcolor[rgb]{0.7, 0, 0}{44.314}} \\
        \({\boldsymbol{I}}_{\boldsymbol{45}}\) & 38.149 & 43.031 & 43.823 & 43.784 & 43.397 & 44.970 & 45.107 & \textbf{\textcolor[rgb]{0, 0, 0.7}{45.323}} & \textbf{\textcolor[rgb]{0, 0.7, 0}{45.345}} & \textbf{\textcolor[rgb]{0.7, 0, 0}{46.079}} \\
        \({\boldsymbol{I}}_{\boldsymbol{90}}\) & 38.197 & 43.243 & 43.875 & 43.988 & 43.384 & \textbf{\textcolor[rgb]{0, 0, 0.7}{45.171}} & 45.121 & 44.907 & \textbf{\textcolor[rgb]{0, 0.7, 0}{45.259}} & \textbf{\textcolor[rgb]{0.7, 0, 0}{45.870}} \\
        \({\boldsymbol{I}}_{\boldsymbol{135}}\) & 37.543 & 42.270 & 43.021 & 42.968 & 42.589 & 44.395 & 44.610 & \textbf{\textcolor[rgb]{0, 0.7, 0}{45.206}} & \textbf{\textcolor[rgb]{0, 0, 0.7}{44.958}} & \textbf{\textcolor[rgb]{0.7, 0, 0}{45.705}} \\
        \({\boldsymbol{S}}_{\boldsymbol{0}}\) & 40.332 & 44.727 & 45.711 & 45.673 & 45.130 & 46.911 & 47.033 & \textbf{\textcolor[rgb]{0, 0.7, 0}{48.254}} & \textbf{\textcolor[rgb]{0, 0, 0.7}{47.900}} & \textbf{\textcolor[rgb]{0.7, 0, 0}{48.406}} \\
        \textbf{\(\textit{DoLP}\)} & 32.748 & 38.113 & 38.849 & 39.009 & 38.748 & 39.743 & \textbf{\textcolor[rgb]{0, 0.7, 0}{39.880}} & 38.806 & \textbf{\textcolor[rgb]{0, 0, 0.7}{39.822}} & \textbf{\textcolor[rgb]{0.7, 0, 0}{40.033}} \\
        \textbf{\(\textit{AoLP}\)} & 23.207 & 26.144 & 25.990 & 26.319 & 26.198 & 26.750 & 27.035 & \textbf{\textcolor[rgb]{0, 0, 0.7}{27.159}} & \textbf{\textcolor[rgb]{0, 0.7, 0}{27.232}} & \textbf{\textcolor[rgb]{0.7, 0, 0}{27.410}} \\
    \bottomrule[2pt]
    \end{tabular}
    \label{tab:psnr}
\end{table}

\begin{table}[htbp]
    \scriptsize
    \centering
    \caption{\textbf{Average RMSE on Simulated Images}}
    \begin{tabular}{c | c | c | c | c | c | c | c | c | c | c}
    \toprule[2pt]
        \textbf{RMSE} & \textbf{NN} & \textbf{BI} & \textbf{BCB} & \textbf{ICPC} & \textbf{EARI} & \textbf{PCDP} & \textbf{PDEC} & \textbf{NP} & \textbf{\uline{LEPD}} & \textbf{\uline{LEIC}} \\
        \midrule[1pt]
        \({\boldsymbol{I}}_{\boldsymbol{0}}\) & 3.2020 & 1.8842 & 1.7533 & 1.7498 & 1.8218 & 1.4969 & 1.4827 & \textbf{\textcolor[rgb]{0, 0, 0.7}{1.4340}} & \textbf{\textcolor[rgb]{0, 0.7, 0}{1.4243}} & \textbf{\textcolor[rgb]{0.7, 0, 0}{1.3281}} \\
        \({\boldsymbol{I}}_{\boldsymbol{45}}\) & 2.8468 & 1.6157 & 1.4871 & 1.4722 & 1.5593 & 1.2400 & 1.2173 & \textbf{\textcolor[rgb]{0, 0, 0.7}{1.1957}} & \textbf{\textcolor[rgb]{0, 0.7, 0}{1.1827}} & \textbf{\textcolor[rgb]{0.7, 0, 0}{1.0878}} \\
        \({\boldsymbol{I}}_{\boldsymbol{90}}\) & 2.8597 & 1.5664 & 1.4802 & 1.4270 & 1.5672 & \textbf{\textcolor[rgb]{0, 0, 0.7}{1.1988}} & 1.2021 & 1.2569 & \textbf{\textcolor[rgb]{0, 0.7, 0}{1.1869}} & \textbf{\textcolor[rgb]{0.7, 0, 0}{1.1102}} \\
        \({\boldsymbol{I}}_{\boldsymbol{135}}\) & 3.0967 & 1.8040 & 1.7024 & 1.6782 & 1.7767 & 1.4198 & 1.3886 & \textbf{\textcolor[rgb]{0, 0.7, 0}{1.3145}} & \textbf{\textcolor[rgb]{0, 0, 0.7}{1.3321}} & \textbf{\textcolor[rgb]{0.7, 0, 0}{1.2406}} \\
        \({\boldsymbol{S}}_{\boldsymbol{0}}\) & 2.0540 & 1.2287 & 1.1338 & 1.1053 & 1.2057 & 0.9386 & 0.9267 & \textbf{\textcolor[rgb]{0, 0.7, 0}{0.8043}} & \textbf{\textcolor[rgb]{0, 0, 0.7}{0.8325}} & \textbf{\textcolor[rgb]{0.7, 0, 0}{0.7945}} \\
        \textbf{\(\textit{DoLP}\)} & 0.0266 & 0.0146 & 0.0135 & 0.0132 & 0.0134 & 0.0122 & \textbf{\textcolor[rgb]{0.7, 0, 0}{0.0120}} & 0.0145 & \textbf{\textcolor[rgb]{0, 0, 0.7}{0.0121}} & \textbf{\textcolor[rgb]{0, 0.7, 0}{0.0121}} \\
        \textbf{\(\textit{AoLP}\)} & 0.0798 & 0.0621 & 0.0608 & 0.0619 & 0.0591 & 0.0582 & 0.0565 & \textbf{\textcolor[rgb]{0, 0.7, 0}{0.0545}} & \textbf{\textcolor[rgb]{0, 0, 0.7}{0.0550}} & \textbf{\textcolor[rgb]{0.7, 0, 0}{0.0538}} \\
    \bottomrule[2pt]
    \end{tabular}
    \label{tab:rmse}
\end{table}

\begin{table}[htbp]
    \scriptsize
    \centering
    \caption{\textbf{Average SSIM on Simulated Images}}
    {
    \begin{tabular}{c | c | c | c | c | c | c | c | c | c | c}
    \toprule[2pt]
        \textbf{SSIM} & \textbf{NN} & \textbf{BI} & \textbf{BCB} & \textbf{ICPC} & \textbf{EARI} & \textbf{PCDP} & \textbf{PDEC} & \textbf{NP} & \textbf{\uline{LEPD}} & \textbf{\uline{LEIC}} \\
        \midrule[1pt]
        \({\boldsymbol{I}}_{\boldsymbol{0}}\) & 0.5921 & 0.5921 & 0.5933 & 0.5724 & 0.6130 & 0.6598 & 0.6606 & \textbf{\textcolor[rgb]{0, 0.7, 0}{0.7158}} & \textbf{\textcolor[rgb]{0, 0, 0.7}{0.7120}} & \textbf{\textcolor[rgb]{0.7, 0, 0}{0.7303}} \\
        \({\boldsymbol{I}}_{\boldsymbol{45}}\) & 0.5842 & 0.5842 & 0.5858 & 0.5650 & 0.6118 & 0.6509 & 0.6569 & \textbf{\textcolor[rgb]{0, 0.7, 0}{0.7116}} & \textbf{\textcolor[rgb]{0, 0, 0.7}{0.7089}} & \textbf{\textcolor[rgb]{0.7, 0, 0}{0.7256}} \\
        \({\boldsymbol{I}}_{\boldsymbol{90}}\) & 0.5819 & 0.5819 & 0.5828 & 0.5630 & 0.6123 & 0.6465 & 0.6470 & \textbf{\textcolor[rgb]{0, 0, 0.7}{0.6936}} & \textbf{\textcolor[rgb]{0, 0.7, 0}{0.6970}} & \textbf{\textcolor[rgb]{0.7, 0, 0}{0.7087}} \\
        \({\boldsymbol{I}}_{\boldsymbol{135}}\) & 0.5874 & 0.5874 & 0.5917 & 0.5676 & 0.6125 & 0.6566 & 0.6590 & \textbf{\textcolor[rgb]{0, 0.7, 0}{0.7174}} & \textbf{\textcolor[rgb]{0, 0, 0.7}{0.7113}} & \textbf{\textcolor[rgb]{0.7, 0, 0}{0.7310}} \\
        \({\boldsymbol{S}}_{\boldsymbol{0}}\) & 0.6657 & 0.6657 & 0.6768 & 0.6515 & 0.6983 & 0.7415 & 0.7386 & \textbf{\textcolor[rgb]{0, 0.7, 0}{0.8115}} & \textbf{\textcolor[rgb]{0, 0, 0.7}{0.8066}} & \textbf{\textcolor[rgb]{0.7, 0, 0}{0.8148}} \\
        \textbf{\(\textit{DoLP}\)} & 0.9294 & 0.9294 & 0.9309 & 0.9322 & 0.9342 & 0.9391 & \textbf{\textcolor[rgb]{0.7, 0, 0}{0.9413}} & 0.9298 & \textbf{\textcolor[rgb]{0, 0.7, 0}{0.9400}} & \textbf{\textcolor[rgb]{0, 0, 0.7}{0.9392}} \\
        \textbf{\(\textit{AoLP}\)} & 0.6861 & 0.6861 & 0.6786 & 0.6778 & 0.6918 & 0.7039 & 0.7193 & \textbf{\textcolor[rgb]{0, 0.7, 0}{0.7378}} & \textbf{\textcolor[rgb]{0, 0, 0.7}{0.7318}} & \textbf{\textcolor[rgb]{0.7, 0, 0}{0.7382}} \\
    \bottomrule[2pt]
    \end{tabular}
    }   
    \label{tab:ssim}
\end{table}

\begin{figure}[htbp]
    \centering
    \includegraphics[width=0.7\linewidth]{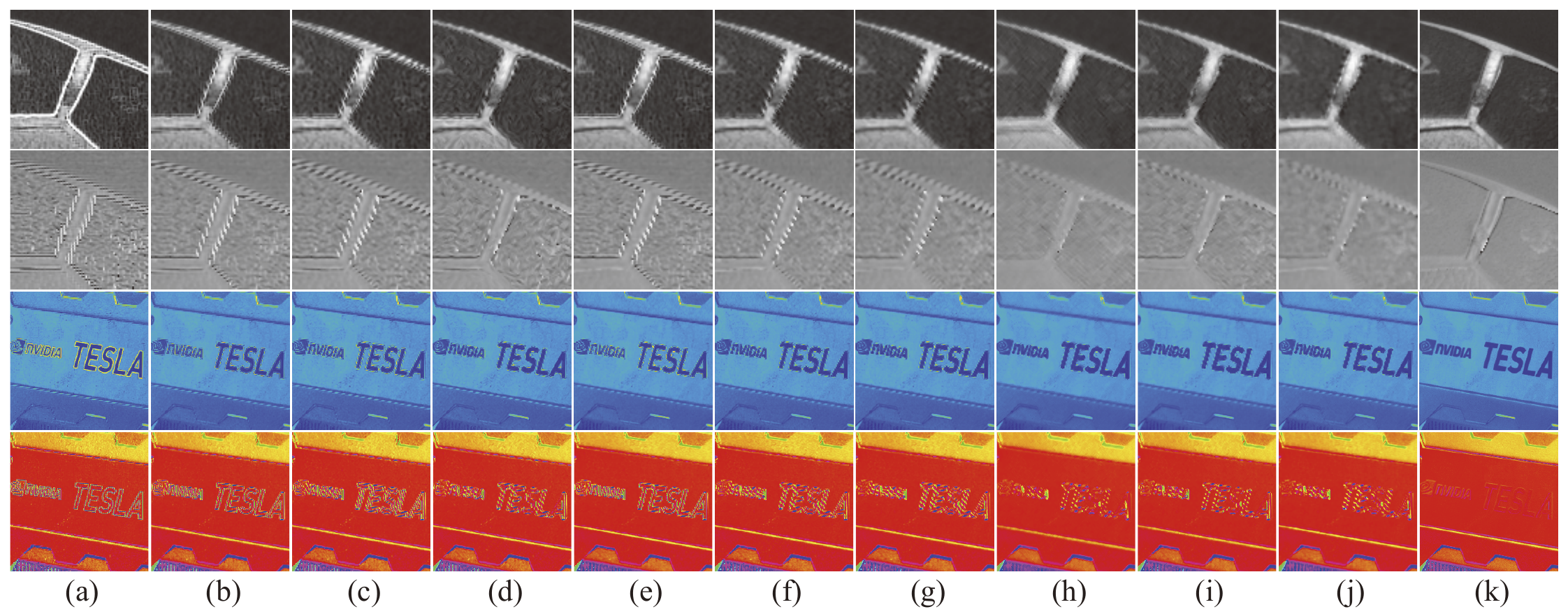}
    \caption{Visual comparison of polarization image reconstruction using different demosaicking techniques. From left to right, the reconstruction results are from methods (a) NN, (b) BI, (c) BCB, (d) ICPC, (e) EARI, (f) PCDP, (g) PDEC, (h) NP, (i) LEPD, (j) LEIC, and (k) ground-truth, respectively.}
    \label{fig:visual_simulated_images}
\end{figure}

Next, we evaluate the demosaicking performance by comparing interpolated images with actual images, as shown in Fig.~\ref{fig:visual_simulated_images}. These images are acquired from the green boxed sections of Fig.~\ref{fig:simulated_dofp_scenes}. For enhanced visual comparisons, DoLP and AoLP images were pseudo-colored using parula and HSV color maps, respectively, in scene 2. Compared with other methods, our methods and NP effectively suppress zipper artifacts around high-frequency, highlighting the importance of the directionality in interpolation decisions. However, NP heavily relies heavily on orthogonal interpolation, which tends to exaggerate textures, introduce artifactual x-like patterns, and reduce detail sharpness, as illustrated in Fig.~\ref{fig:visual_simulated_images}(h). 
In contrast, our LEPD and LEIC methods exhibit significant refinements on these issues. Moreover, compared with LEPD, LEIC employs ICCC to enhance fidelity by filtering out certain redundant interpolated textures, further improving the quality and clarity of the images.

Finally, the running times of the various demosaicking methods are evaluated, with the results presented in Table~\ref{tab:running_time_simulated_images}. Notably, traditional methods such as NN, BI, and BCB are excluded from the speed comparisons due to their unsatisfactory performance quality. NP demonstrates robust performance in previous tests but incurs significant computational costs due to its complex interpolation decision-making process and the larger computation window required. In contrast, our LEPD method exhibits the fastest speed, being 26\% faster than the second-fastest method, PCDP. Our LEIC method builds upon LEPD and incorporates ICCC, which adds certain additional time costs but achieves the best overall interpolation performance. The running time of LEIC is comparable to that of PDEC and is significantly lower than both NP and EARI by 92.46\% and 388.90\%, respectively. This efficiency in processing time, coupled with superior image quality, highlights the effectiveness of our LEIC method in balancing performance with computational efficiency.

\begin{table}[htbp]
    \scriptsize
    \centering
    \caption{\textbf{Running Time on Simulated DoFP Images}}
    {
    \begin{tabular}{c | c | c | c | c | c | c | c}
    \toprule[2pt]
        \textbf{Method} & \textbf{ICPC} & \textbf{EARI} & \textbf{PCDP} & \textbf{PDEC} & \textbf{NP} & \textbf{\uline{LEPD}} & \textbf{\uline{LEIC}} \\
        \midrule[1pt]
        \textbf{Time Cost (\textit{s})} & 0.4648 & 1.3166 & \textbf{\textcolor[rgb]{0, 0.7, 0}{0.1895}} & \textbf{\textcolor[rgb]{0, 0, 0.7}{0.2610}} & 0.5186 & \textbf{\textcolor[rgb]{0.7, 0, 0}{0.1402}} & 0.2693 \\
    \bottomrule[2pt]
    \end{tabular}
    }   
    \label{tab:running_time_simulated_images}
\end{table}

\subsection{Test on real DoFP images}
\begin{figure}[htbp]
    \centering
    \includegraphics[width=0.6\linewidth]{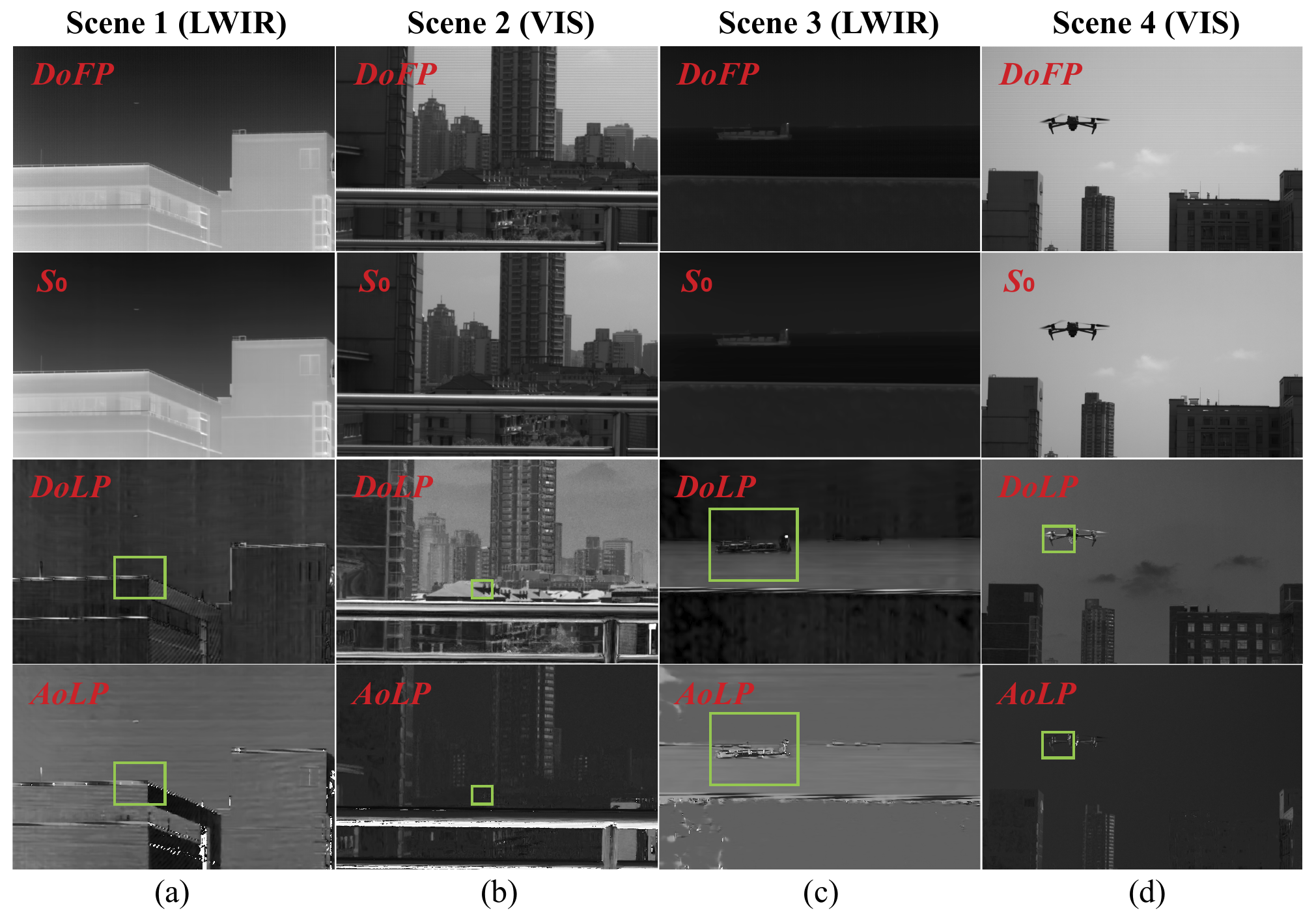}
    \caption{Four real DoFP scenes of OLVD; certain scenes include distant targets such as ships or UAVs. Rows 1 to 4 display DoFP, \({{S}_{0}}\), $\textit{DoLP}$, and $\textit{AoLP}$ images, respectively.}
    \label{fig:authentic_dofp_scenes}
\end{figure}

In this experiment, we selected four real DoFP images from our OLVD\cite{Liu2024OLVD} for testing, as displayed in Fig.~\ref{fig:authentic_dofp_scenes}. Scenes 1 and 3 were captured using the North Guangwei UMC4A-PU0A Micro DoFP LWIR polarization imager. This device incorporates an array of wire-grid micro-polarizers\cite{yongqiang2018infrared}, features a resolution of 640 $\times$ 512, and records images with a depth of 14 bits. Scenes 2 and 4 were captured using the Daheng Imaging MER2-503-36U3M POL DoFP visible polarization imager, which utilizes a monochromatic quad-polarizer array\cite{sswpolarization}. This imager operates at a resolution of 2448 $\times$ 2048, an 8-bit depth, and a frame rate of 36 fps. Additionally, the pBM3D method\cite{abubakar2018block} was employed to denoise the DoFP images, ensuring the enhancement of image quality by reducing noise while preserving important polarization information.

\begin{figure} [htbp]
    \centering
    \includegraphics[width=0.7\linewidth]{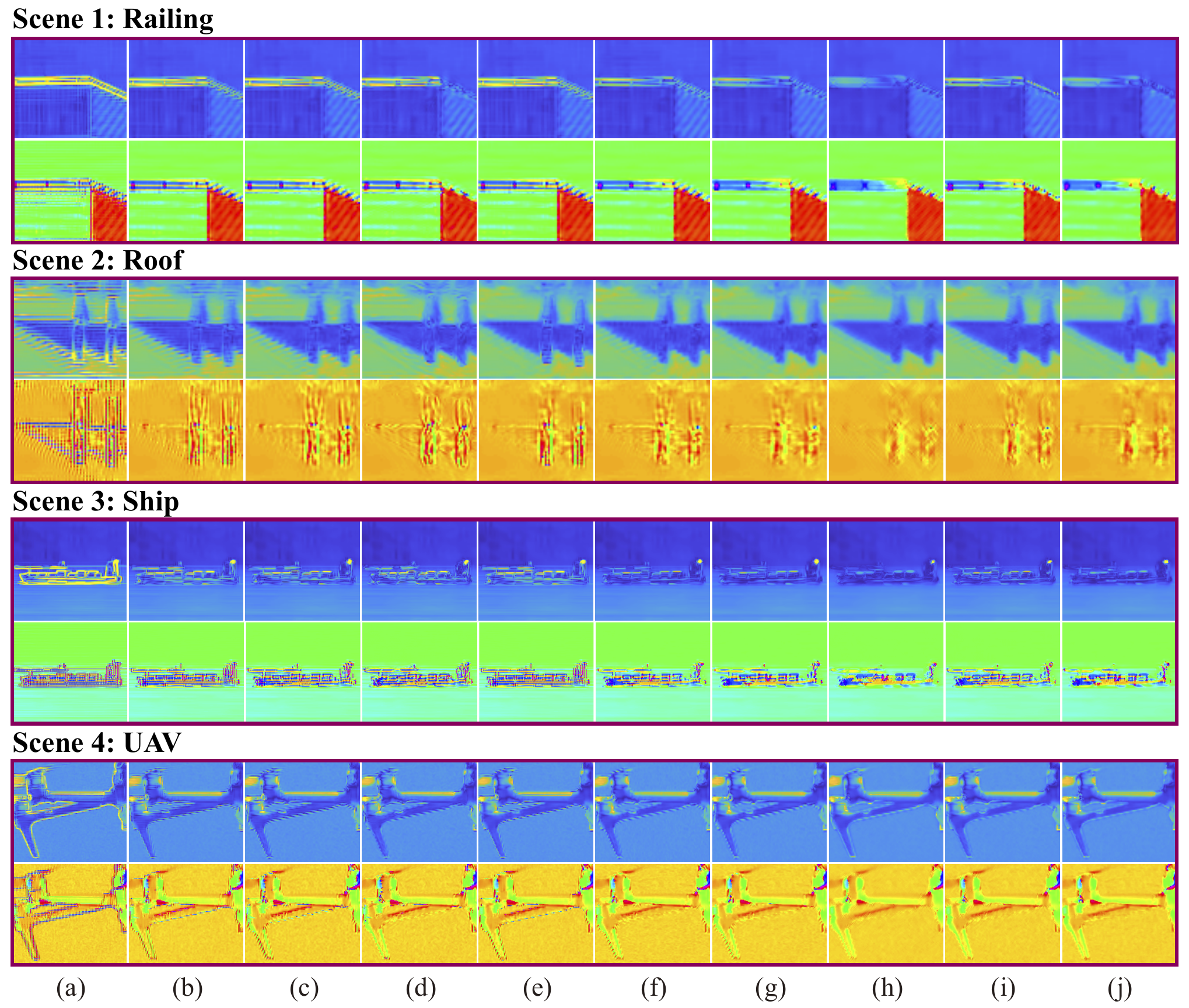}
    \caption{Visual comparison of demosaicking methods on real DoFP images. The first and second rows in each scene correspond to $\textit{DoLP}$ and $\textit{AoLP}$, respectively. From left to right, the reconstruction results are presented by methods (a) NN, (b) BI, (c) BCB, (d) ICPC, (e) EARI, (f) PCDP, (g) PDEC, (h) NP, (i) our LEPD and (j) LEIC. }
    \label{fig:visual_authentic_images}
\end{figure}

Fig.~\ref{fig:visual_authentic_images} illustrates the $\textit{DoLP}$ and $\textit{AoLP}$ results using various interpolation methods for the green boxed areas of Fig.~\ref{fig:authentic_dofp_scenes}. For enhanced visual comparison, similar to Fig.~\ref{fig:synthesis_process}, the $\textit{DoLP}$ and $\textit{AoLP}$ images are processed in parula and HSV pseudo-colors, respectively. 

Traditional demosaicking methods, as shown in Fig.~\ref{fig:visual_authentic_images}(a), (b), and (c), produce significant artifacts in both LWIR and VIS DoFP images. Advanced methods achieve varying degrees of artifact mitigation. Edge-aware-based methods such as ICPC, EARI, and NP reduce false edges to some extent. However, they share a common issue of injecting redundant false textures due to the instability of the interpolation decision maker, as demonstrated in Fig.~\ref{fig:visual_authentic_images}(d), (e), and (h). A typical example includes the x-like textures produced by NP, notably visible in scenes 1 and 2 of Fig.~\ref{fig:visual_authentic_images}(h). PCDP and PDEC reduce incorrect high-frequency information through multi-channel difference domain interpolation; however, the absence of directional interpolation decisions often leads to significant artifacts, as evident in Fig.~\ref{fig:visual_authentic_images}(f) and (g).

In contrast, LEPD matches or even surpasses the performance of PCDP and PDEC in artifact treatment, such as those observed around the UAV in scene 4 of Fig.~\ref{fig:visual_authentic_images}(i). By employing ICCC, LEIC further diminishes artifacts and false textures. LEIC’s performance is comparable to that of NP but with minimal additional textural errors. In summary, LEIC significantly enhances the fidelity and visualization of reconstructed polarization images.

Finally, Table~\ref{tab:time_real} lists the runtimes of various high-precision demosaicking methods on the DoFP images of LWIR and VIS. Among them, LEPD is the fastest, while LEIC offers the best reconstruction performance with satisfactory computational costs, balancing efficiency with effectiveness in processing DoFP images.

\begin{table}[htbp]
    \scriptsize
    \centering
    \caption{\textbf{Running Time on Real DoFP Images}}
    {
    \begin{tabular}{c | c | c | c | c | c | c | c}
    \toprule[2pt]
        \textbf{Time Cost (\textit{s})} & \textbf{ICPC} & \textbf{EARI} & \textbf{PCDP} & \textbf{PDEC} & \textbf{NP} & \textbf{\uline{LEPD}} & \textbf{\uline{LEIC}} \\
        \midrule[1pt]
        \textbf{\(\text{LWIR}\)} & 0.1393 & 0.3229 & \textbf{\textcolor[rgb]{0, 0.7, 0}{0.0629}} & \textbf{\textcolor[rgb]{0, 0, 0.7}{0.0807}} & 0.1512 & \textbf{\textcolor[rgb]{0.7, 0, 0}{0.0405}} & 0.0935 \\
        \textbf{\(\text{VIS}\)} & 3.1531 & 9.0719 & \textbf{\textcolor[rgb]{0, 0.7, 0}{1.1519}} & \textbf{\textcolor[rgb]{0, 0, 0.7}{1.5329}} & 2.7124 & \textbf{\textcolor[rgb]{0.7, 0, 0}{0.8014}} & 1.7440 \\
    \bottomrule[2pt]
    \end{tabular}
    }   
    \label{tab:time_real}
\end{table}

\section{Conclusions}
We introduce an efficient LEIC method for demosaicking images captured by DoFP sensors, known as LEIC. This method is based on a standardized framework called the TIPDF. LEIC incorporates a low-cost edge-aware strategy, termed DLE, that guides the interpolation process using a Logistic function to account for differences in directional variations. Additionally, LEIC utilizes an ICCC technique to refine the demosaicked results. For applications requiring a less resource-intensive solution, we propose LEPD as a lightweight version of LEIC, which omits the ICCC component. Among 10 tested interpolation-based demosaicking methods, LEIC emerged as superior in quantitative metrics and visual effects while maintaining low computational costs. Meanwhile, LEPD ranks as the swiftest among the non-traditional methods. In scenarios where accuracy is prioritized over speed, LEPD can serve as a rapid, high-quality initializer for more complex DoFP demosaicking programs. Owing to their high parallelizability, our methods can recover high-resolution images in real-time on GPUs or FPGAs, rendering them highly suitable for industrial applications.

\bibliographystyle{IEEEtran}
\bibliography{IEEEabrv,aaai25}	

\begin{thebibliography}{10}
\providecommand{\url}[1]{#1}
\csname url@samestyle\endcsname
\providecommand{\newblock}{\relax}
\providecommand{\bibinfo}[2]{#2}
\providecommand{\BIBentrySTDinterwordspacing}{\spaceskip=0pt\relax}
\providecommand{\BIBentryALTinterwordstretchfactor}{4}
\providecommand{\BIBentryALTinterwordspacing}{\spaceskip=\fontdimen2\font plus
\BIBentryALTinterwordstretchfactor\fontdimen3\font minus \fontdimen4\font\relax}
\providecommand{\BIBforeignlanguage}[2]{{%
\expandafter\ifx\csname l@#1\endcsname\relax
\typeout{** WARNING: IEEEtran.bst: No hyphenation pattern has been}%
\typeout{** loaded for the language `#1'. Using the pattern for}%
\typeout{** the default language instead.}%
\else
\language=\csname l@#1\endcsname
\fi
#2}}
\providecommand{\BIBdecl}{\relax}
\BIBdecl

\bibitem{kechiche2020polarimetric}
A.~Z. Kechiche, O.~Aubreton, A.~Mathieu, A.~Mannucci, and C.~Stolz, ``Polarimetric imaging method for surface quality evaluation of a liquid metal pool obtained during welding,'' \emph{Optical Engineering}, vol.~59, no.~10, pp. 100\,501--100\,501, 2020.

\bibitem{li2021near}
X.~Li, F.~Liu, P.~Han, S.~Zhang, and X.~Shao, ``Near-infrared monocular 3d computational polarization imaging of surfaces exhibiting nonuniform reflectance,'' \emph{Optics Express}, vol.~29, no.~10, pp. 15\,616--15\,630, 2021.

\bibitem{usmani2021deep}
K.~Usmani, G.~Krishnan, T.~O’Connor, and B.~Javidi, ``Deep learning polarimetric three-dimensional integral imaging object recognition in adverse environmental conditions,'' \emph{Optics Express}, vol.~29, no.~8, pp. 12\,215--12\,228, 2021.

\bibitem{li2021illumination}
N.~Li, Y.~Zhao, Q.~Pan, S.~G. Kong, and J.~C.-W. Chan, ``Illumination-invariant road detection and tracking using lwir polarization characteristics,'' \emph{ISPRS Journal of Photogrammetry and Remote Sensing}, vol. 180, pp. 357--369, 2021.

\bibitem{romano2012day}
J.~M. Romano, D.~Rosario, and J.~McCarthy, ``Day/night polarimetric anomaly detection using spice imagery,'' \emph{IEEE Transactions on Geoscience and Remote Sensing}, vol.~50, no.~12, pp. 5014--5023, 2012.

\bibitem{zhu2019depth}
D.~Zhu and W.~A. Smith, ``Depth from a polarisation+ rgb stereo pair,'' in \emph{Proceedings of the IEEE/CVF conference on computer vision and pattern recognition}, 2019, pp. 7586--7595.

\bibitem{hu2016polarization}
F.~Hu, Y.~Cheng, L.~Gui, L.~Wu, X.~Zhang, X.~Peng, and J.~Su, ``Polarization-based material classification technique using passive millimeter-wave polarimetric imagery,'' \emph{Applied optics}, vol.~55, no.~31, pp. 8690--8697, 2016.

\bibitem{liang2020effective}
Z.~Liang, X.~Ding, Z.~Mi, Y.~Wang, and X.~Fu, ``Effective polarization-based image dehazing with regularization constraint,'' \emph{IEEE Geoscience and Remote Sensing Letters}, vol.~19, pp. 1--5, 2020.

\bibitem{garcia2018bio}
M.~Garcia, C.~Edmiston, T.~York, R.~Marinov, S.~Mondal, N.~Zhu, G.~P. Sudlow, W.~J. Akers, J.~Margenthaler, S.~Achilefu \emph{et~al.}, ``Bio-inspired imager improves sensitivity in near-infrared fluorescence image-guided surgery,'' \emph{Optica}, vol.~5, no.~4, pp. 413--422, 2018.

\bibitem{niu2018low}
Y.~Niu, J.~Ouyang, W.~Zuo, and F.~Wang, ``Low cost edge sensing for high quality demosaicking,'' \emph{IEEE transactions on image processing}, vol.~28, no.~5, pp. 2415--2427, 2018.

\bibitem{wu2016bayer}
J.~Wu, M.~Anisetti, W.~Wu, E.~Damiani, and G.~Jeon, ``Bayer demosaicking with polynomial interpolation,'' \emph{IEEE Transactions on Image Processing}, vol.~25, no.~11, pp. 5369--5382, 2016.

\bibitem{yang2020mcfd}
X.~Yang, W.~Zhou, and H.~Li, ``Mcfd: A hardware-efficient noniterative multicue fusion demosaicing algorithm,'' \emph{IEEE Transactions on Circuits and Systems for Video Technology}, vol.~31, no.~9, pp. 3575--3589, 2020.

\bibitem{lien2017efficient}
C.-Y. Lien, F.-J. Yang, and P.-Y. Chen, ``An efficient edge-based technique for color filter array demosaicking,'' \emph{IEEE sensors journal}, vol.~17, no.~13, pp. 4067--4074, 2017.

\bibitem{menon2006demosaicing}
D.~Menon, S.~Andriani, and G.~Calvagno, ``Demosaicing with directional filtering and a posteriori decision,'' \emph{IEEE Transactions on Image Processing}, vol.~16, no.~1, pp. 132--141, 2006.

\bibitem{kiku2016beyond}
D.~Kiku, Y.~Monno, M.~Tanaka, and M.~Okutomi, ``Beyond color difference: Residual interpolation for color image demosaicking,'' \emph{IEEE Transactions on Image Processing}, vol.~25, no.~3, pp. 1288--1300, 2016.

\bibitem{mihoubi2018survey}
S.~Mihoubi, P.-J. Lapray, and L.~Bigu{\'e}, ``Survey of demosaicking methods for polarization filter array images,'' \emph{Sensors}, vol.~18, no.~11, p. 3688, 2018.

\bibitem{ratliff2009interpolation}
B.~M. Ratliff, C.~F. LaCasse, and J.~S. Tyo, ``Interpolation strategies for reducing ifov artifacts in microgrid polarimeter imagery,'' \emph{Optics express}, vol.~17, no.~11, pp. 9112--9125, 2009.

\bibitem{gao2011bilinear}
S.~Gao and V.~Gruev, ``Bilinear and bicubic interpolation methods for division of focal plane polarimeters,'' \emph{Optics express}, vol.~19, no.~27, pp. 26\,161--26\,173, 2011.

\bibitem{zhang2016image}
J.~Zhang, H.~Luo, B.~Hui, and Z.~Chang, ``Image interpolation for division of focal plane polarimeters with intensity correlation,'' \emph{Optics express}, vol.~24, no.~18, pp. 20\,799--20\,807, 2016.

\bibitem{li2019demosaicking}
N.~Li, Y.~Zhao, Q.~Pan, and S.~G. Kong, ``Demosaicking dofp images using newton’s polynomial interpolation and polarization difference model,'' \emph{Optics express}, vol.~27, no.~2, pp. 1376--1391, 2019.

\bibitem{morimatsu2020monochrome}
M.~Morimatsu, Y.~Monno, M.~Tanaka, and M.~Okutomi, ``Monochrome and color polarization demosaicking using edge-aware residual interpolation,'' in \emph{2020 IEEE International Conference on Image Processing (ICIP)}.\hskip 1em plus 0.5em minus 0.4em\relax IEEE, 2020, pp. 2571--2575.

\bibitem{wu2021polarization}
R.~Wu, Y.~Zhao, N.~Li, and S.~G. Kong, ``Polarization image demosaicking using polarization channel difference prior,'' \emph{Optics Express}, vol.~29, no.~14, pp. 22\,066--22\,079, 2021.

\bibitem{xin2023demosaicking}
J.~Xin, Z.~Li, S.~Wu, and S.~Wang, ``Demosaicking dofp images using edge compensation method based on correlation,'' \emph{Optics Express}, vol.~31, no.~9, pp. 13\,536--13\,551, 2023.

\bibitem{zhang2018sparse}
J.~Zhang, H.~Luo, R.~Liang, A.~Ahmed, X.~Zhang, B.~Hui, and Z.~Chang, ``Sparse representation-based demosaicing method for microgrid polarimeter imagery,'' \emph{Optics letters}, vol.~43, no.~14, pp. 3265--3268, 2018.

\bibitem{li2023joint}
N.~Li, B.~Wang, F.~Goudail, Y.~Zhao, and Q.~Pan, ``Joint denoising-demosaicking network for long-wave infrared division-of-focal-plane polarization images with mixed noise level estimation,'' \emph{IEEE Transactions on Image Processing}, 2023.

\bibitem{zeng2019end}
X.~Zeng, Y.~Luo, X.~Zhao, and W.~Ye, ``An end-to-end fully-convolutional neural network for division of focal plane sensors to reconstruct s 0, dolp, and aop,'' \emph{Optics express}, vol.~27, no.~6, pp. 8566--8577, 2019.

\bibitem{pistellato2022deep}
M.~Pistellato, F.~Bergamasco, T.~Fatima, and A.~Torsello, ``Deep demosaicing for polarimetric filter array cameras,'' \emph{IEEE Transactions on Image Processing}, vol.~31, pp. 2017--2026, 2022.

\bibitem{hamilton1997adaptive}
J.~F. Hamilton~Jr, ``Adaptive color plan interpolation in signal sensor color electronic camera,'' \emph{United State Patent, 5,629,734}, 1997.

\bibitem{buades2009self}
A.~Buades, B.~Coll, J.-M. Morel, and C.~Sbert, ``Self-similarity driven color demosaicking,'' \emph{IEEE Transactions on Image Processing}, vol.~18, no.~6, pp. 1192--1202, 2009.

\bibitem{gharbi2016deep}
M.~Gharbi, G.~Chaurasia, S.~Paris, and F.~Durand, ``Deep joint demosaicking and denoising,'' \emph{ACM Transactions on Graphics (ToG)}, vol.~35, no.~6, pp. 1--12, 2016.

\bibitem{qiu2019polarization}
S.~Qiu, Q.~Fu, C.~Wang, and W.~Heidrich, ``Polarization demosaicking for monochrome and color polarization focal plane arrays,'' in \emph{International Symposium on Vision, Modeling and Visualization, 2019}.\hskip 1em plus 0.5em minus 0.4em\relax The Eurographics Association, 2019.

\bibitem{sun2019hybrid}
B.~Sun, N.~Yuan, and Z.~Zhao, ``A hybrid demosaicking algorithm for area scan industrial camera based on fuzzy edge strength and residual interpolation,'' \emph{IEEE Transactions on Industrial Informatics}, vol.~16, no.~6, pp. 4038--4048, 2019.

\bibitem{li2008image}
X.~Li, B.~Gunturk, and L.~Zhang, ``Image demosaicing: A systematic survey,'' in \emph{Visual Communications and Image Processing 2008}, vol. 6822.\hskip 1em plus 0.5em minus 0.4em\relax SPIE, 2008, pp. 489--503.

\bibitem{chung2006color}
K.-H. Chung and Y.-H. Chan, ``Color demosaicing using variance of color differences,'' \emph{IEEE transactions on image processing}, vol.~15, no.~10, pp. 2944--2955, 2006.

\bibitem{Liu2024OLVD}
L.~Guangsen, ``{OLVD dataset},'' \url{https://github.com/lgs195/OLVD}.

\bibitem{yongqiang2018infrared}
Z.~Yongqiang, L.~Ning, Z.~Peng, Y.~Jiaxin, and P.~Quan, ``Infrared polarization perception and intelligent processing,'' \emph{Infrared and Laser Engineering}, vol.~47, no.~11, p. 1102001, 2018.

\bibitem{sswpolarization}
S.~Group \emph{et~al.}, ``Polarization image sensor with four-directional on-chip polarizer and global shutter function.''

\bibitem{abubakar2018block}
A.~Abubakar, X.~Zhao, S.~Li, M.~Takruri, E.~Bastaki, and A.~Bermak, ``A block-matching and 3-d filtering algorithm for gaussian noise in dofp polarization images,'' \emph{IEEE Sensors Journal}, vol.~18, no.~18, pp. 7429--7435, 2018.

\end{thebibliography}

\end{document}